\documentclass[reprint,amsmath,amssymb,nofootinbib,aps,prc]{revtex4-1}
\usepackage{graphicx}
\usepackage{multirow}
\usepackage{epstopdf}
\usepackage{caption}
\usepackage{color}
\usepackage{bm}
\usepackage{textcomp} 
 
\begin{document}

\title{Charge asymmetry dependence of flow and a novel correlator to detect the chiral magnetic wave in a multiphase transport model}

\author{Diyu Shen$^{a,b}$} \thanks{shendiyu@sinap.ac.cn}
\author{Jinhui Chen$^{c}$} \thanks{chenjinhui@fudan.edu.cn}
\author{Guoliang Ma$^{c}$}
\author{Yu-Gang Ma$^{a,c}$}
\author{Qiye Shou$^{c}$} \thanks{shouqiye@fudan.edu.cn}
\author{Song Zhang$^{c}$}
\author{Chen Zhong$^{c}$}
\affiliation{$^a$Shanghai Institute of Applied Physics, Chinese Academy of Sciences, Shanghai 201800, China}
\affiliation{$^b$University of Chinese Academy of Sciences, Beijing 100049, China}
\affiliation{$^c$Key Laboratory of Nuclear Physics and Ion-beam Application (MOE), Institute of Modern Physics, Fudan University, Shanghai 200433, China}

\begin{abstract}
In a multiphase transport model with the initial electric quadrupole moment, we studied and discussed the charge asymmetry ($A_{\rm ch}$) dependence of flow at varied kinematic windows in semi-central Au+Au collisions at $\sqrt{s_{\rm NN}}$ = 200 GeV. We then proposed a novel correlator $W$ which specially focuses on the difference of elliptic flow between positively and negatively charged hadrons induced by the chiral magnetic wave and, more importantly, is irrelevant to the ambiguous $A_{\rm ch}$. We found that the distribution of the second order correlator $W_2$ displays a convex structure in the absence of the quadrupole and a concave shape in the presence of the quadrupole. We then studied the response of $W_n$ for both signal and resonance background in a toy model and in analytical calculation. Such a method provides a new way to detect the chiral magnetic wave.
\end{abstract}

\maketitle

\section{Introduction} \label{sec:intro}

It is theorized that the chiral anomaly is able to give rise to the imbalance between the numbers of left- and right-handed quarks in quantum chromodynamics (QCD) matter \cite{Kharzeev2008,Kharzeev2015}. Such a novel phenomenon, originally stemming from the topological structure of QCD vacuum as well as the possible local $\cal P$ and/or $\cal CP$ violation in the strong interaction, has drawn extensive attentions over the past decades \cite{Kharzeev2016}. In the presence of an external strong magnetic field ($\overrightarrow{B}$), the interplay of the chiral anomaly and the magnetic field is proposed to generate more intriguing anomalous chiral phenomena, e.g. the chiral magnetic effect (CME) \cite{Fukushima2008} and the chiral separation effect (CSE) \cite{Liao2010}. The CME is expected to induce an electric current along the direction of $\overrightarrow{B}$ while the CSE leading to a chirality current according to the relation: $\overrightarrow{J_{e(5)}} \propto \mu_{5(e)}\overrightarrow{B}$, where subscripts $5$ and $e$ denote the chirality and electric term respectively. Realistically, relativistic heavy-ion collisions, by which the Quark-Gluon Plasma (QGP) is expected to be produced \cite{reviewQGP1,reviewQGP2}, provide a unique environment to experimentally investigate such anomalous chiral effects~\cite{reviewNST}. For the case of CME, a finite electric dipole moment with respect to the reaction plane could be formed and, with the help of the appropriate observables, is feasible to be detected \cite{Voloshin2004}. Based on this idea, plenty of experimental analyses have been carrying out at RHIC and LHC to test such charge separations \cite{Zhao2019a}.

Furthermore, the chiral magnetic wave (CMW), a collective excitation of the CME and CSE, has also been theoretically proposed \cite{Burnier2011,Burnier2012,Yee2014}. By analogy with the dipole moment arising from the CME, the CMW could manifest itself in forming an electric quadrupole moment of the collision system, where the ``poles" (out-of-plane) and the ``equator" (in-plane) of the participant region respectively acquire additional positive or negative charges, depending on the net charge of the system. It is suggested that such a CMW-induced charge separation can be examined via the charge asymmetry ($A_{\rm ch}$) dependence of elliptic flow ($v_2$), the second-order Fourier component of the particle azimuthal distribution \cite{Poskanzer1998}, between the positive and negatively charged particles, i.e.,
\begin{equation} \label{eq:1}
v_{2}^{\pm} - v_{\rm 2, base}^{\pm} = \mp \frac{a}{2}A_{\rm ch},
\end{equation}
where superscript $\pm$ denotes the charge of the particles, $v_{2,\rm base}$ represents the ``usual" $v_{2}$ unrelated to the charge separation, $a$ is the quadrupole moment normalized by the net charge density and $A_{\rm ch} = (N^{+} -N^{-}) / (N^{+} +N^{-})$ with $N$ denoting the number of particles measured in a given event, or more concisely,
\begin{equation} \label{eq:2}
\Delta v_{2} \equiv v_{2}^{-} - v_{2}^{+} \simeq rA_{\rm ch},
\end{equation}
where the slope parameter $r$ is used to quantify the strength of the quadrupole configuration.
\begin{figure*}
\captionsetup{justification=raggedright}
\includegraphics[width=32pc]{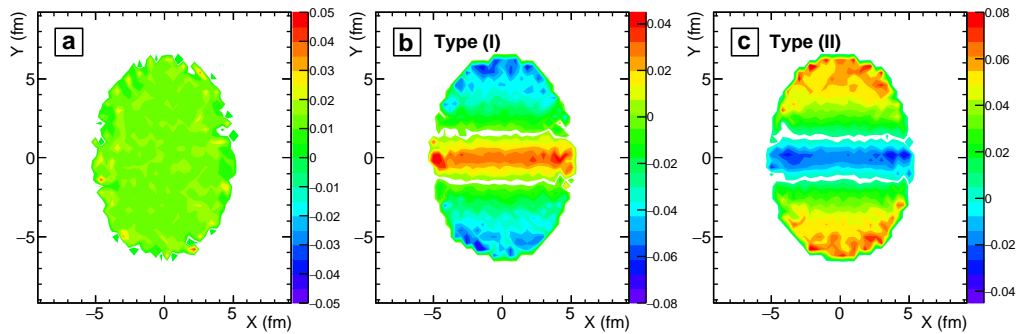}
\captionof{figure}{The net electric charge distributions of partons in the transverse plane for 30 - 40\% central Au+Au collisions at $\sqrt{s_{\rm NN}}$ = 200 GeV when the initial electric quadrupole moment is (a) not imported; (b) imported for the events with $A_{\rm ch}<-0.01$; (c) imported for the events with $A_{\rm ch}>-0.01$ in AMPT model.}
\label{fig:quadrupole}
\end{figure*}

In recent years, the CMW has attracted wide theoretical attentions \cite{Stephanov2013,Han2019,Zhao2019b} and the slope $r$ has also been measured in experiment. The results of the experimental search for the CMW have been reported by the ALICE \cite{Voloshin2014,Adam2016}, CMS \cite{Park2017,Sirunyan2017} and STAR \cite{Adamczyk2015,Shou2019} collaborations at various collision energies and systems. In the heavy ion collisions, i.e., Au+Au, U+U collisions at RHIC and Pb+Pb collisions at LHC, the extracted slopes in the semi-central collisions, e.g., 30 - 40\% most central, are basically consistent with each other and are of the same order of magnitude as predicted by theory \cite{Burnier2011} regardless of the collision energy scales, seemingly supporting the CMW expectation. On the other hand, the very similar slope has been surprisingly observed in p+Pb collisions, which indicates the existence of a common background since the direction of $\overrightarrow{B}$ is decoupled from the reaction plane and the chiral anomaly is not expected to be created in the small system collisions \cite{Belmont2017}. The main background effects of the CMW measurement, according to the theoretical estimations, include the local charge conservation \cite{Bzdak2013}, the baryon stopping \cite{Campbell2013} and the isospin chemical potential \cite{Hatta2016} entwined with the collectivity of the QGP. All of these effects are likely to contribute to a certain part, if not all, of the experimental observations, hence the existence of the CMW still remains inconclusive.

It is important to stress that both $A_{\rm ch}$ and $v_2$ in Eqs. (\ref{eq:1}) and (\ref{eq:2}) are experimental observables and the slope $r$ can only be correctly extracted when $A_{\rm ch}$ and $v_2$ are calculated in the appropriate phase space. It is well known that $v_2$ considerably depends on the transverse momentum ($p_{\rm T}$) and pseudorapidity ($\eta$), therefore it is worthwhile to investigate the dependence of $v_2$ on $A_{\rm ch}$ in varied kinematic windows. The charge asymmetry $A_{\rm ch}$ is constructed to characterize, by definition, the net charge of the whole collision system, nevertheless, the experimental measurement of $A_{\rm ch}$ strongly depends on the detector acceptance and tracking efficiency, as pointed out in Refs.~\cite{Voloshin2014,Adam2016}. As a result, a small uncertainty of the ambiguous $A_{\rm ch}$ could lead to the wrong estimation of the slope especially considering that the signal of $r$ is small ($\approx$1\%). To get rid of this issue, it is noteworthy to develop new methods which can accurately capture the CMW-induced electric quadrupole.

In this work we perform a study of the $A_{\rm ch}$ dependence of flow and then propose a novel, $A_{\rm ch}$-free correlator to detect the CMW in Au+Au collisions at $\sqrt{s_{\rm NN}}$ = 200 GeV with the modified AMPT model, in which the initial electric quadrupole configuration is designedly imported. This paper is organized as follows. In Sec. \ref{sec:method}, we give a brief introduction to the model and the methodology used in our study. The dependence of $v_2$ on $A_{\rm ch}$ in varied kinematic windows is discussed in Sec. \ref{sec:v2ach}. The new correlator $W(\Delta v_2)$ is introduced and presented in Sec. \ref{sec:w}. The effect from resonance decay contribution to the new correlator is also studied with independent model and analytical calculation in Sec. \ref{sec:wbg}. A summary and conclusions are given in Sec. \ref{sec:sum}.

\section{Model and method} \label{sec:method}

The AMPT model is a hybrid transport model widely used in the study of the high-energy heavy-ion collisions. In particular, the string melting version used in this analysis is known for the success in describing the collective behavior of the final state hadrons \cite{Lin2005}. The model consists of four subroutines which simulate, in sequence, different stages of the evolution in the collisions. The initial parton conditions are generated by HIJING \cite{Wang1991}. The evolution of the partonic phase is performed by the Zhang's Parton Cascade (ZPC) model \cite{Zhang1998}, and then a hadronization process handled by the quark coalescence is implemented to form hadrons. The rescatterings and interactions of the hadronic matter are processed in the ART model. For this work, we set the parameters of Lund string fragmentation function $a$ and $b$ to be 2.2 and 0.5 respectively, and the cross section of parton scattering to be 10 mb with a Debye screening mass $\mu=$1.767/fm, so that the hadron spectrum and anisotropic flow at RHIC energy can be reasonably reproduced. 

In order to mimic the electric quadrupole moment generated by CMW, we adopt the approach proposed in \cite{Ma2011,Ma2014}, which interchanges the $y$ component of the position coordinate for some in-plane light quarks carrying positive (negative) charges with those out-plane ones carrying negative (positive) charges at the beginning of the partonic stage. Such an operation gives rise to two cases of charge separations: (\uppercase\expandafter{\romannumeral1}) for those events with the negative net charge ($A_{\rm ch}<-0.01$)$\footnote{The value 0.01 comes from a tiny but positive intercept reported by STAR \cite{Adamczyk2015}.}$, a percentage of the positive quarks, i.e. $u$ and $\bar{d}$, are set to be concentrated on the ``equator" while $\bar{d}$ and $u$ gathering around the ``poles", as shown in Fig.~\ref{fig:quadrupole} (b); (\uppercase\expandafter{\romannumeral2}) the opposite configuration is applied for the events with the positive net charge ($A_{\rm ch}>-0.01$), as shown in Fig.~\ref{fig:quadrupole} (c). The strength of the quadrupole moment is tuned by such percentages, and we refer them as Type (I), Type (II) configuration, respectively. The distribution from original AMPT without quadrupole moment is also present in Fig.~\ref{fig:quadrupole} (a) for reference. According to the previous study, switching 2-3\% of quarks could generate a comparable CMW signal as predicted by the theory and preliminarily observed in semi-central collisions in the experiment. A relatively stronger value 10\% is set for the study of the novel correlator, which will be discussed in Sec. \ref{sec:w}.

\begin{figure}
\captionsetup{justification=raggedright}
\includegraphics[width=16pc]{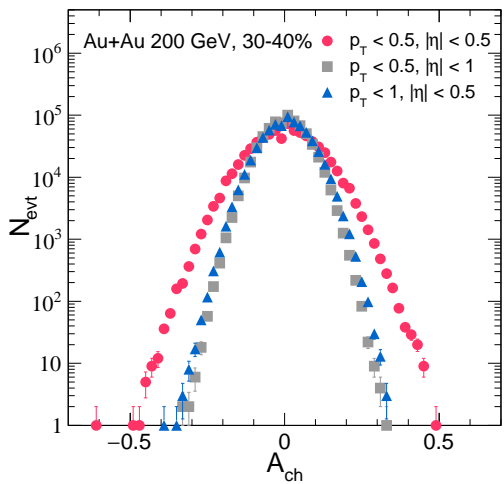}
\captionof{figure}{The distribution of charge asymmetry for 30 - 40\% central Au+Au $\sqrt{s_{\rm NN}}$ = 200 GeV collisions at different phase space windows in AMPT model.}
\label{fig:ach}
\end{figure}
The charge asymmetry is calculated by all charged hadrons in a given kinematic window and it forms a symmetric distribution with the mean value locating around zero$\footnote{In fact, the mean value is slightly above zero with the order of $10^{-3}$.}$, as shown in Fig. \ref{fig:ach}. It is found that the width of the distribution becomes narrower as the multiplicity increases due to the weaker fluctuation. For the flow calculation, the event plane method \cite{Poskanzer1998} is used and the result is checked to be well consistent with the experiment result. All hadrons are selected in $|\eta|<1$ to imitate the detector acceptance unless otherwise noted.

\section{Dependence of $v_2$ on $A_{\rm ch}$ in varied kinematic windows}  \label{sec:v2ach}

The linear dependence of $v_2$ for $h^\pm$ on $A_{\rm ch}$ is commonly taken as a possible signal of CMW with the slope $r$ characterizing its strength. It is noticed that, however, the $r$ extracted from the linear fit between $\Delta v_2$ and $A_{\rm ch}$ can be influenced by a few factors, e.g., the experimental correction of the observed $A_{\rm ch}$ owing to the limited detecting ability and, more significantly, the kinematic windows employed when measuring $A_{\rm ch}$ and flow. It is well known that $v_2$ has a remarkable dependence on the transverse momentum. The varied average values of the transverse momentum ($\langle p_{\rm T} \rangle$) in different $A_{\rm ch}$ intervals inevitably generate a trivial non-zero slope. The $A_{\rm ch}$ dependence of $\langle p_{\rm T} \rangle$ has been examined in our work. It is found that there is no indication of such a mechanism for the primordial hadrons, namely, $\langle p_{\rm T} \rangle$ remains unchanged regardless of $A_{\rm ch}$ at any $p_{\rm T}$ coverage before the hadronic interaction. It is also confirmed that the hadronic scatterings play negligible roles in this study since they neither produce extra $v_2$ splitting nor severely change $A_{\rm ch}$. Hence, we infer that the most likely source of the $\langle p_{\rm T} \rangle$-$A_{\rm ch}$ correlation as observed in the CMS data \cite{Sirunyan2017} is something beyond the processes mentioned above, such as the LCC.

\begin{figure}
\captionsetup{justification=raggedright}
\includegraphics[width=20pc]{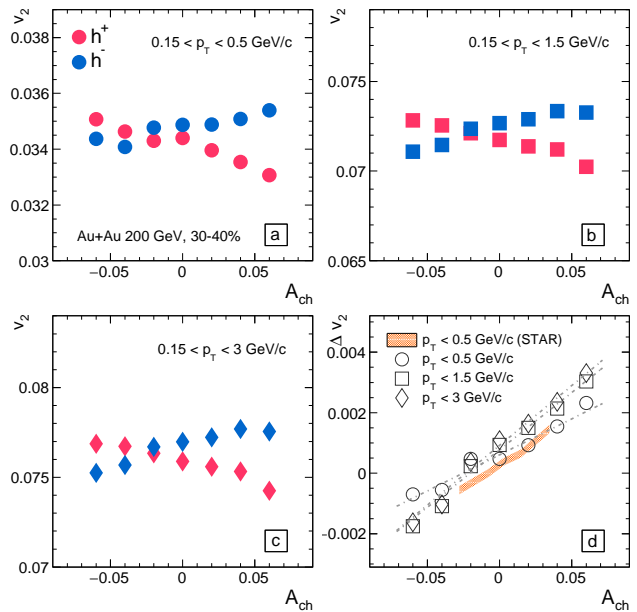}
\captionof{figure}{$v_2$ and $\Delta v_{2}$ for $h^\pm$ in different $p_{\rm T}$ windows as functions of $A_{\rm ch}$ for 30 - 40\% central Au+Au collisions at $\sqrt{s_{\rm NN}}$ = 200 GeV in AMPT model. The dashed lines in panel (d) represent the linear fits to the given results in AMPT model, while the orange band is experimental data from STAR Collaboration~\cite{Adamczyk2015}.}
\label{fig:v2}
\end{figure}

The dependences of $v_2$ on $A_{\rm ch}$ in three $p_{\rm T}$ intervals are presented in Fig.~\ref{fig:v2}, for 30 - 40\% central Au+Au collisions at $\sqrt{s_{\rm NN}}$ = 200 GeV. The lower limit of $p_{\rm T}$ is set to be 0.15 GeV/$c$ for the consistence with the experiment. In all three cases, the $v_2$ for $h^+$ decrease while the $v_2$ for $h^-$ increase as the increase of $A_{\rm ch}$. The linear relationship between $\Delta v_2$ and $A_{\rm ch}$ can also be clearly observed. The extracted $r$ in Fig.~\ref{fig:v2} (d) are 2.4\%, 3.8\% and 3.9\% for $p_{\rm T} < 0.5$ GeV/$c$, $p_{\rm T} < 1.5$ GeV/$c$ and  $p_{\rm T} < 3$ GeV/$c$ respectively, which perfectly reproduce the STAR data of $\sim$3\% \cite{Adamczyk2015}. In our model, particles regardless of $p_{\rm T}$ are all influenced by the imported charge separation, so the slope can be obtained at both the narrow and the wide $p_{\rm T}$ windows. We suggest to perform such a study in the experiment. If the measured slopes at various $p_{\rm T}$ windows keep constant, it can be partially, if not all, attributed to the CMW. The STAR collaboration has reported that there is little change of the slopes between the $p_{\rm T}$ coverage of $< 0.5$ GeV/$c$ and $< 2$ GeV/$c$~\cite{Shou2019} at RHIC energy scales, seemingly matching our model study. We now encourage to perform such a measurement at the LHC energy scales and at the small system collisions to check the component of the signal and the backgrounds. The CMS measurement was only carried out for $p_{\rm T} < 3$ GeV/$c$. If the slope rapidly reduces or even vanishes when the $p_{\rm T}$ windows gets narrower, one can deduce that the trivial $\langle p_{\rm T} \rangle$-$A_{\rm ch}$ correlation plays a dominated role. 

In addition to the $p_{\rm T}$ coverage, we also reduced the $\eta$ coverage to half unity to check the variation of $r$.  In the presence of the robust CMW signal in our model, the slope stays unchanged when the $\eta$ window varies because of the negligible dependence of $v_2$ on $\eta$ at mid-rapidity. On the other hand, the LCC predicts a notable difference as $\eta$ varies~\cite{Bzdak2013}. Such features should also be examined in the experiments to draw further conclusions.

\begin{figure}
\captionsetup{justification=raggedright}
\includegraphics[width=20pc]{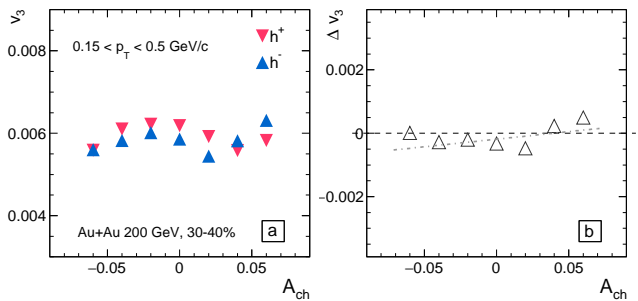}
\captionof{figure}{$v_3$ and $\Delta v_{3}$ for $h^\pm$ as a function of $ A_{\rm ch}$ for 30 - 40\% central Au+Au collisions at $\sqrt{s_{\rm NN}}$ = 200 GeV in AMPT model. The dashed lines represent the linear fits to the given data points.}
\label{fig:v3}
\end{figure}

The $A_{\rm ch}$ dependent $v_3$ is considered as a valid reference for detecting the CMW, since the quadrupole is not supposed to establish any relation between $v_3$ and $A_{\rm ch}$. Fig.~\ref{fig:v3} depicts the $v_3$ and $\Delta v_3$ for $h^\pm$ as a function of $A_{\rm ch}$. Not surprisingly, no linear dependence is found for $v_3$ on $A_{\rm ch}$ in the presence of the imported electric quadrupole. An extracted $r$ of $\sim0.4\%$ demonstrates the smallness of the trivial backgrounds in the model and the reliability for $v_3$ serving as a baseline. Therefore, the observed non-zero $v_3$ slopes as reported in Refs.~\cite{Sirunyan2017,Shou2019} indicate the existence of the non-CMW background effects.

One should be aware that the $A_{\rm ch}$ should be measured as accurately as possible, so that the net charge of the whole collision system can be truthfully characterized. Modifying the cuts for $A_{\rm ch}$ must cause the change of $r$ because when calculating $A_{\rm ch}$ in a nonuniform phase space, one is preferentially selecting one of the two quadrupole configurations, i.e., either Type (I) or Type (II) of Fig.~\ref{fig:quadrupole}. The linear dependence of $\Delta v_2$ on $A_{\rm ch}$ is then masked. The experimental measurement of $A_{\rm ch}$ strongly depends on the detecting condition, which makes the search for the CMW more ambiguous. For this concern, Refs.~\cite{Voloshin2014,Adam2016} have suggested a method which is free of $A_{\rm ch}$ correction by measuring the covariance instead of the slope between $\Delta v_2$ and $A_{\rm ch}$. In this work, we further suggest a novel correlator which is $A_{\rm ch}$-free to detect the CMW-generated electric quadrupole moment.

\section{W Correlator} \label{sec:w}

It was proposed in Refs.~\cite{Ajitanand2011,Magdy2018a,Magdy2018b} that a charge-sensitive in-event correlator is able to effectively test the CME by measuring and comparing the $sine$ term of the expanded azimuthal distribution for negatively and positively charged hadrons in the real and shuffled events. And experimental data analysis on this correlator is progressing well~\cite{YeQM18}. Given the fact that the difference between CME and CMW measurements, in essence, is nothing but the former focusing on the non-zero $sine$ term while the latter aiming at the extra $cosine$ contribution, i.e., $v_2$, it is feasible to extend such a correlator to detect the electric quadrupole moment besides the dipole. Based on this idea, we first construct a distribution in the real event,
\begin{equation} \label{eq:r}
N(\Delta v_2)^{\rm real} = \frac{\sum_{1}^{p}cos(2\Delta\phi^+)}{p} - \frac{\sum_{1}^{n}cos(2\Delta\phi^-)}{n},
\end{equation}
where $p$ and $n$ are the numbers of positive and negative hadrons respectively, and $\Delta\phi^{+(-)}$ is the azimuthal emission angle of $h^{+(-)}$ with respect to event plane. As its name implies, this real distribution is obtained from the factual event and is used to probe the possible charge dependence of elliptic flow. Meanwhile, we analogously construct a shuffled distribution serving as the reference,

\begin{figure}
\captionsetup{justification=raggedright}
\includegraphics[width=20pc]{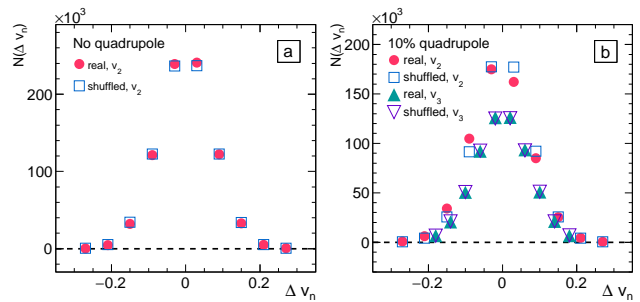}
\captionof{figure}{The distribution of $N(\Delta v_2^{\rm real})$ and $N(\Delta v_2^{\rm shuf.})$ (a) without and (b) with the initial quadrupole moment for 30 - 40\% central Au+Au collisions at $\sqrt{s_{\rm NN}}$ = 200 GeV in AMPT model.}
\label{fig:nDistr}
\end{figure}

\begin{figure*}
\captionsetup{justification=raggedright}
\includegraphics[width=42pc]{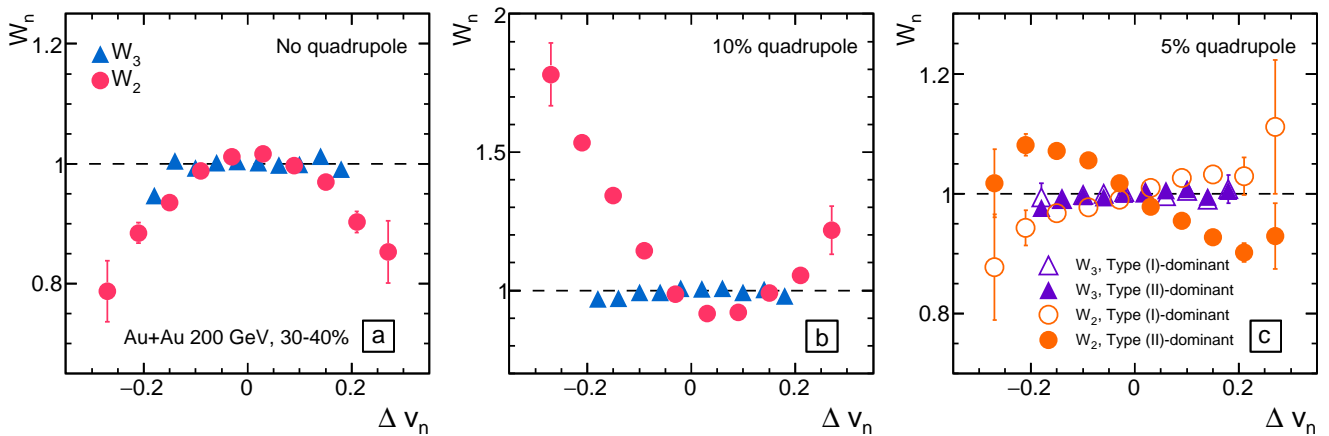}
\captionof{figure}{The correlator $W_{2(3)}$ as a function of $\Delta v_{2(3)}$ (a) without and (b) with the initial quadrupole moment in AMPT model. Panel (c) represents the results for different quadrupole moment type in Fig.~\ref{fig:quadrupole} (b) or (c). }
\label{fig:w2w3}
\end{figure*}

\begin{equation} \label{eq:s}
N(\Delta v_2)^{\rm shuf.} = \frac{\sum_{1}^{p}cos(2\Delta\phi)}{p} - \frac{\sum_{1}^{n}cos(2\Delta\phi)}{n},
\end{equation}
in which $p$ and $n$ keep the same numbers while the charge of interest hadron is re-shuffled (randomly selected). These two distributions can also be extended to the higher order harmonic flow, e.g., $v_3$. Ideally, if each particle is distributed independently with a specific probability density function (PDF) of $\phi$ one would expect the distribution of $v_2$ of numerator for real events will be border than denominator for shuffled events. Because CMW brings an extra event by event fluctuations of $v_2$, while the distribution of $v_3$ for different charge particles remain the same and therefor have no difference between numerator and denominator. On the other hand, as point out by Ref. \cite{YC Feng2018}, the correlation between opposite charge particles due to resonance decay will bring additional contribution to the final results.

The distributions of Eqs. (\ref{eq:r}) and (\ref{eq:s}) are presented and compared in Fig.~\ref{fig:nDistr}. The panel (a) shows the real and shuffled distributions as a function of $\Delta v_{2}$ without any quadrupole configuration, while panel (b) shows the case of $v_2$ and $v_3$ with 10\% quadrupole being imported. For $v_2$, it can be observed that the real and shuffled distributions behave differently when there's a finite quadrupole, however, the discrepancy decreases as the quadrupole goes to zero. For $v_3$, two distributions stay the same even when the charged separation is implemented. Such trends are consistent with what one expects. In order to closely evaluate the configuration and magnitude of the electric quadrupole, we define a new correlator which is the ratio of the above two distributions,
\begin{equation} \label{eq:w}
W_{n}(\Delta v_n) = \frac{N({\Delta v_n})^{\rm real}}{N({\Delta v_n})^{\rm shuf.}},
\end{equation}
where $n$ is chosen to be 2 or 3. In doing so, general collective properties other than charge separation are ensured to be identical for the numerator and the denominator.

Figure~\ref{fig:w2w3} presents the correlator $W_n$ calculated by the charged hadrons with $0.15 < p_{\rm T} < 1.5$ GeV/$c$ and $|\eta|<1$ as a function of $\Delta v_n$ in 30 - 40\% central Au+Au collisions at $\sqrt{s_{\rm NN}}$ = 200 GeV. The panel (a) and (b) depict the behavior for $W$ without and with initial quadrupole. In the absence of the initial charge separation, a convex structure of $W_2$ can be clearly observed with $W_{2} \approx$1 at $\Delta v_2=0$ and $W_{2}<1$ as $\Delta v_2$ increases and decreases. Such a convex shape is consistent with the detection of CME with $R(\Delta S)$ correlator as reported in \cite{Ajitanand2011} and serves as a baseline of the background effect. When the initial quadrupole is implemented, the convex shape turns out to be varied. When the quadrupole is set to 10\%, the $W_2$ displays a notable concave distribution with the value larger than 1 at the non-zero $\Delta v_2$ and slightly smaller than 1 at $\Delta v_2=0$. Unlike the symmetric convex in Fig.~\ref{fig:w2w3} (a) with the minimum at $\Delta v_2=0$, the concave in Fig.~\ref{fig:w2w3} (b) proves to be asymmetric with the minimum at a positive $\Delta v_2$, which can be interpreted by the domination of the events with a specific quadrupole configuration. For a better understanding of the structure of $W$, we set the percentage at 5\% and preferentially select those events with different quadrupole moment type as shown in Fig.~\ref{fig:w2w3} (c). It can be clearly seen that a positive correlation is formed between $W_2$ and $\Delta v_2$ for the event sample dominated by Type (I) configuration, and to the contrary, an inverse correlation is formed for the event sample dominated by Type (II). Since the data sample collected in the experiment is always a mixture of the two quadrupole configurations as shown in Fig.~\ref{fig:quadrupole} if the CMW exists, the minimum of the $W_2$ could be helpful to determine the content and ratio of such two configurations. Compared with $W_2$, the charge separation has very limited effect on $W_3$ because of the irrelevant orientation of third order event plane to the magnetic field. The calculated $W_3$ values are consistent with unity regardless of $\Delta v_3$ with and without quadrupole, indicating the smallness of the background effects in the model and the applicability of $W_3$ serving as a reference.

\begin{figure}
\captionsetup{justification=raggedright}
\includegraphics[width=20pc]{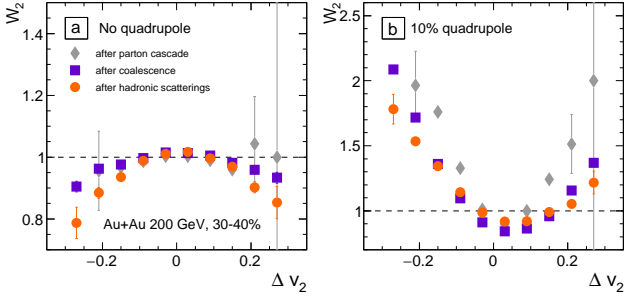}
\captionof{figure}{The evolution of $W_2$($\Delta v_2$) from different stages of the heavy-ion collision (a) without and (b) with the initial quadrupole moment in AMPT model.}
\label{fig:w2evo}
\end{figure}

It is essential to study how different stages of the collision influence the correlator. The time evolution of $W_2$ is presented and compared in Fig.~\ref{fig:w2evo}. The left and right panel show $W_2$ as a function of $\Delta v_2$ without and with quadrupole respectively, at the stages of parton and hardron. In the absence of the charge separation, the $W_2$ exhibits the convex structure after the process of the parton cascade and behaves similarly right after the hadronization with the coalescence, as shown in the diamond and square markers. After the process of hadronic scatterings, however, the $W_2$ distribution turns out to be more convex, which indicates the discernible contribution from the resonance decay. In the presence of the quadrupole, the $W_2$ distribution shows a more concave shape at the partonic state and the curvature gradually decreases at the hadronic state. It can be seen that both the hadronization and the final state interactions play visible roles in the evolution of $W_2$ by smearing the initial charge separation. Since the resonance decays dominate at the low-$p_{\rm T}$ range, we suggest to increase the $p_{\rm T}$ range for the experimental measurement in order to avoid such a background effect.

\section{Analytical calculation and a toy model simulation of background subtraction} \label{sec:wbg}

As we mentioned above, the process of resonance decay could generate correlation between different charge particles, which will bring additional contribution to both the $W_2$ and $W_3$ correlator. Indeed, the convex or concave shape of $W_{\rm n}$ correlator depends on the relative width of the distribution of numerator and denominator. It was suggested that the process of resonance decay could bring considerable background to $R$ correlator as used in CME research~\cite{Piotr2018,YC Feng2018}. From this point of view, we present an analytical calculation and a toy model simulation with resonance decay influence to the shape of $W_{\rm n}$ correlator in this section. This is independent of the above analysis in AMPT model. Unless specified, we assume that the yield of $\pi^+$ is equal to the $\pi^-$ in the following calculation. In the experimental analysis one can select the same amount of $\pi^+$ and $\pi^-$ artificially. For convenience, we use the following notations:
\begin{eqnarray} \label{eq:notation}
c_{n} &=& \frac{\sum\limits_{i=0}\limits^{n}cos(2\Delta\phi_{i})}{n}, \notag \\
\sigma^2 &=& \mathrm{Var}[c_n] 
\end{eqnarray} 
with sum over all particles in an event, thus $c_n$ becomes an event by event fluctuation observable with the variance of $\sigma^2$. 

The $W_2$ correlator is then rewritten as:
\begin{equation} \label{W:rewrite}
W_2 = \frac{c_{n}^{+} - c_{n}^{-}}{c_{n}^{1} - c_{n}^{2}}
\end{equation}
where $c_{\rm n}^{+}$, $c_{\rm n}^{-}$ are calculated with $\pi^+$, $\pi^-$, and $c_{\rm n}^{1}$, $c_{\rm n}^{2}$ are from charge-shuffled pions, respectively. In the case of no CMW effect, the variance of numerator can be expressed as follows,
\begin{eqnarray} \label{Var:numerator}
\sigma^2_{\mathrm{num}} &=& \sigma_{+}^{2} + \sigma_{-}^{2} - 2\mathrm{Cov}(c_{n}^{+}, c_{n}^{-}) \notag \\
&=& 2\sigma^2 - 2\mathrm{Cov}(c_{n}^{+}, c_{n}^{-})
\end{eqnarray}
where $\sigma_{+}^{2} = \sigma_{-}^{2}$ is based on the symmetric consideration of event average. Similarly, we can get the variance of denominator
\begin{equation} \label{Var:denominator}
\sigma^2_{\mathrm{den}} = \sigma_{1}^{2} + \sigma_{2}^{2} - 2\mathrm{Cov}(c_{n}^{1}, c_{n}^{2}) 
\end{equation}
where
\begin{eqnarray}
\sigma_{1}^{2} = \sigma_{2}^{2} 
&=& \frac{\sigma_{+}^{2}}{2} + \frac{\sigma_{-}^{2}}{2} 
+ 2\mathrm{Cov}(c_{n/2}^{+}, c_{n/2}^{-}) \notag \\
&=& \sigma^2 + 2\mathrm{Cov}(c_{n/2}^{+}, c_{n/2}^{-})
\end{eqnarray}
\begin{eqnarray} \label{Cov:denominator}
2\mathrm{Cov}(c_{n}^{1}, c_{n}^{2}) &=& 2(\mathrm{Cov}(c_{n/2}^{1+}, c_{n/2}^{2+}) + \mathrm{Cov}(c_{n/2}^{1-}, c_{n/2}^{2-}) \notag \\
& & + \mathrm{Cov}(c_{n/2}^{1+}, c_{n/2}^{2-}) + \mathrm{Cov}(c_{n/2}^{1-}, c_{n/2}^{2+})) \notag \\
&=& 4\mathrm{Cov}(c_{n/2}^{+}, c_{n/2}^{-})
\end{eqnarray}
Here we use $\sigma_{\rm n}^{2} = 2\sigma_{\rm n/2}^{2}$, which is the central limit theory conclusion. In Eq.(\ref{Cov:denominator}) we assume that the correlation between same charge particles can be neglected because the process of $\rho$ decay always generates correlation between opposite charged particles. Therefore, the ratio of the variances of numerator and denominator of $W_2$ correlator can be expressed as:
\begin{equation} \label{Ratio}
R^2=\frac{\sigma_{\mathrm{num}}^{2}}{\sigma_{\mathrm{den}}^{2}} = \frac{2\sigma^2 - 2\mathrm{Cov}(c_{n}^{+}, c_{n}^{-})}{2\sigma^2}
\end{equation}
So the convex or concave shape of $W_2$ depends on the covariance of different charge particles when there is no CMW effect. If each particle is independently distributed with the same PDF, $W_2$ will be a constant value of unity. In order to make it clear how the covariance be influenced by resonance decay, we only focus on the pair of $\pi^{\pm}$ from a $\rho$ decay. In that case, the covariance can be write as:
\begin{eqnarray} \label{Cov:resonance}
\mathrm{Cov}(c_{n}^{+}, c_{n}^{-}) 
&=& \langle cos(2\Delta\phi^+)cos(2\Delta\phi^-) \rangle \notag \\
& &- \langle cos(2\Delta\phi^+)\rangle \langle cos(2\Delta\phi^-) \rangle \notag \\
&=& \langle \frac{cos(4\Delta\phi_\rho) + cos(2\delta)}{2} \rangle \notag \\
& &- \langle cos(2\Delta\phi_\rho+\delta) \rangle \langle cos(2\Delta\phi_\rho-\delta) \rangle \notag \\
&=&\frac{1}{2} \langle cos(2\delta) \rangle + \frac{v_{4,\rho}}{2} - v_{2,\rho}^{2} \langle cos(\delta) \rangle^2 \notag \\
&\approx&\frac{1}{2} \langle cos(2\delta) \rangle 
\end{eqnarray}
where $\Delta\phi_{\rho} = (\Delta\phi_{+}+\Delta\phi_{-})/2$ is the azimuthal angle of $\rho$ resonance with respect to the reaction plane, the $\delta = \phi_{+} - \phi_{-}$ is the open angle of decay pair. Here we have two assumptions: (1) in a resonance decay, $(\phi_{+} + \phi_{-})/2$ could be treat as the direction of mother resonance, (2) the open angle of decay particles is independent of the $\rho$ direction~\cite{YC Feng2018}. We use $\langle sin\delta \rangle $= 0 in the last step of Eq.(\ref{Cov:resonance}). This is because $\delta$ is symmetrically distributed around zero, and the bracket denotes average over all events. It can be straightforward to get the covariance of an event with $n_{\rm \pi}$ and $n_{\rm \rho}$ by simply multiplying Eq.(\ref{Cov:resonance}) by $n_{\rm \rho}/(0.5n_{\rm \pi})^2$, where $n_{\rm \pi}$ and $n_{\rm \rho}$ are the total number of pions and $\rho$ resonances in an event, respectively. According to the typical value of higher order harmonic flow in Au+Au collisions at $\mathrm{\sqrt{s_{NN}}}$ = 200 GeV~\cite{PHENIX:flow}, one can conclude that the leading order of Eq.(\ref{Cov:resonance}) is $\langle cos(2\delta)/2 \rangle$. It is worth emphasising that we simply assume $v_2$ and $v_4$ sharing the same reaction plane in this calculation. It means that the contribution of $v_4$ to the second order event plane will be further decreased when one consider the correlation between second order event plane and forth order event plane~\cite{PHENIX:flow}. 

The contribution of CMW effect to the shape of $W_n$ correlator can also be analytically calculated. In the following calculation, for convenience, we neglect the covariance term of Eq.(\ref{Ratio}). It can be easily recovered if one take the resonance decay background into consideration. Before going to the $W_n$ correlator, it is helpful to consider the variance of an observable with the following PDF: 
\begin{equation}
f = \frac{1}{2}[f_{1}(\mu_1,\sigma_1^{2}) + f_{2}(\mu_2,\sigma_2^{2})]
\end{equation}
where $f_1$ and $f_2$ are the two sub-distributions with mean value $\mu_1$ and $\mu_2$, variance $\sigma_1$ and $\sigma_2$, respectively. By definition we can calculate the variance of observable $x_i$ as:
\begin{eqnarray}
\sigma^{2} &=& \int x^2f(x)dx - [\int f(x)xdx]^2 \notag \\
&=&\frac{1}{2}[\int x^2f_1(x)dx + \int x^2f_2(x)dx] - \frac{1}{4}(\mu_{1} + \mu_{2})^2 \notag \\
&=&\frac{1}{2}(\sigma_{1}^{2}+\sigma_{2}^{2})+\frac{1}{4}(\mu_1 - \mu_2)^2
\end{eqnarray}
It can be seen that the variance of summed distribution not only depends on the variance of two sub-distribution but also on the difference of their mean value. In order to derive the analytical formulation of $W_n$ correlator to CMW effect, we rewrite Eq.(\ref{Ratio}) as
\begin{eqnarray}\label{Var:cmw}
R^2=\frac{\sigma_{\mathrm{num}}^2}{\sigma_{\mathrm{den}}^2}&=&\frac{\mathrm{Var}[\frac{1}{2}f_1(-\Delta v_2,2\sigma^{2})+\frac{1}{2}f_2(+\Delta v_2,2\sigma^{2})]}{\mathrm{Var}[f(0,\sigma^{2})]} \notag \\
&=&\frac{2\sigma^2 + (\Delta v_2)^2}{2\sigma^2}
\end{eqnarray} 
where $\sigma^2$ can be calculated by central limit theory, i.e.
\begin{eqnarray}\label{Var:sigma_W2}
\sigma^2 &=& \frac{1}{N}(\int cos^2(2\phi)f(\phi)d\phi - [\int cos(2\phi)f(\phi)d\phi]^2) \notag \\
&=& \frac{1}{N}(\frac{1}{2}+\frac{v_4}{2}-v_{2}^{2}) \notag \\
&\approx&\frac{1}{2N}
\end{eqnarray}
Here N is the particle number of interest, $f(\phi)$ is the PDF of particle's azimuthal angle with respect to the reaction plane.
This analysis can be easily performed to $W_3$ correlator by replacing the $cos(2\Delta\phi)$ with $cos(3\Delta\phi)$, i.e.
\begin{eqnarray} \label{Cov:resonance(W3)}
\mathrm{Cov}_{(W_3)}&=&\frac{1}{2} \langle cos(3\delta) \rangle + \frac{v_{6,\rho}}{2} - v_{3,\rho}^{2} \langle cos(\frac{3\delta}{2}) \rangle^2 \notag \\
&\approx&\frac{1}{2} \langle cos(3\delta) \rangle
\end{eqnarray}
\begin{eqnarray}\label{Var:sigma_W3}
\sigma^2_{(W_3)} &=& \frac{1}{N}(\int cos^2(3\phi)f(\phi)d\phi - [\int cos(3\phi)f(\phi)d\phi]^2) \notag \\
&=& \frac{1}{N}(\frac{1}{2}+\frac{v_6}{2}-v_{3}^{2}) \notag \\
&\approx&\frac{1}{2N}
\end{eqnarray}

In the toy model simulation, we modify the azimuthal distribution of different charge primordial pions to imitate the signal of CMW and vary kinematic properties of $\rho$ resonance to study the background, ie.
\begin{equation} \label{eq:pdf-pi}
f_{\pi}(\Delta\phi^{\pm}) = 1 + 2(v_{2,\pi}^{\mathrm{def}} \mp a )cos(2\Delta\phi^{\pm}) + 2v_{3,\pi}cos(3\Delta\phi^{\pm})
\end{equation}
and
\begin{equation} \label{eq:pdf-rho}
f_{\rho}(\Delta\phi) = 1 + 2v_{2,\rho}cos(2\Delta\phi) + 2v_{3,\rho}cos(3\Delta\phi)
\end{equation}
where $f_{\pi}(\Delta\phi^{\pm})$ and $f_{\rho}(\Delta\phi)$ are the PDF of azimuthal angle distribution with respect to the reaction plane of $\pi^{\pm}$ and $\rho$ resonance, respectively, $v_{2,\pi}^{\mathrm{def}}$ and $v_{2,\rho}$ are the default elliptic flow spectra as described by Refs.~\cite{FQ Wang2017, Aihong2019}, the sign of $\Delta v_2$ fluctuates from event to event with same probability. The triangle flow $v_3$ of both primordial pions and $\rho$ resonance at any given kinematic window is set to be 1/5 of the corresponding $v_2^{\mathrm{def}}~$\cite{Qiye2014}. Meanwhile, $v_1$ and other higher order flow harmonics are neglected according to Eq.(\ref{Cov:resonance}) and Eq.(\ref{Cov:resonance(W3)}). In this simulation, we build the event with 390 charged pions in it (195 for each charged type). In the scenario with resonance decay contribution effect, it could be consisted of 324 primordial charge pions, and 33 $\rho$ resonances that each decays into a $\pi^+$ + $\pi^-$ pair~\cite{Aihong2019}. This configuration gives a total multiplicity that matches the multiplicity within 2 units of rapidity for $30-40\%$ central Au+Au collisions at $\mathrm{\sqrt{s_{NN}}}$ = 200 GeV~\cite{STAR2009}.

\begin{figure}[!htb]
\centering
\includegraphics[width=0.48\linewidth]{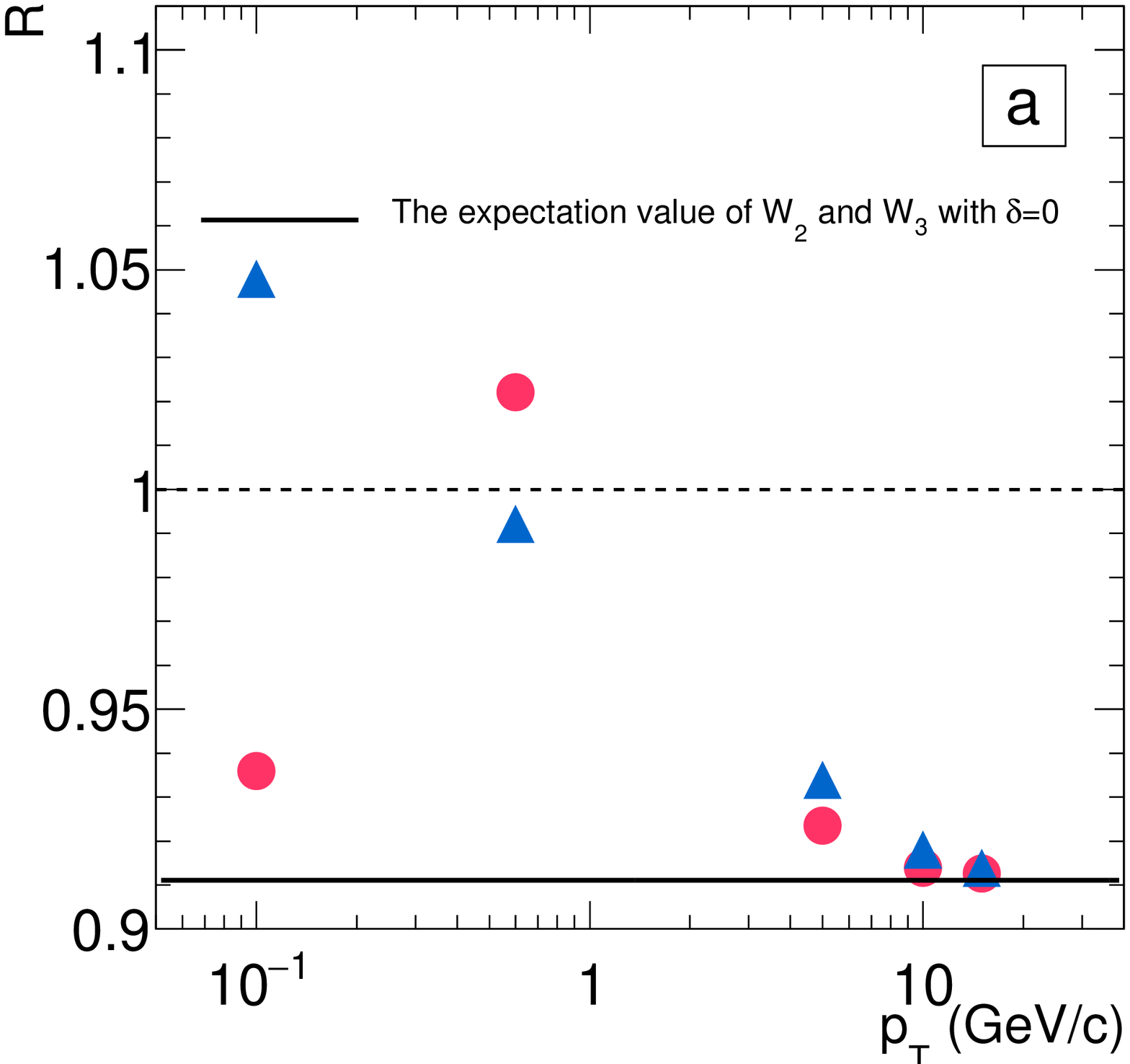} 
\includegraphics[width=0.48\linewidth]{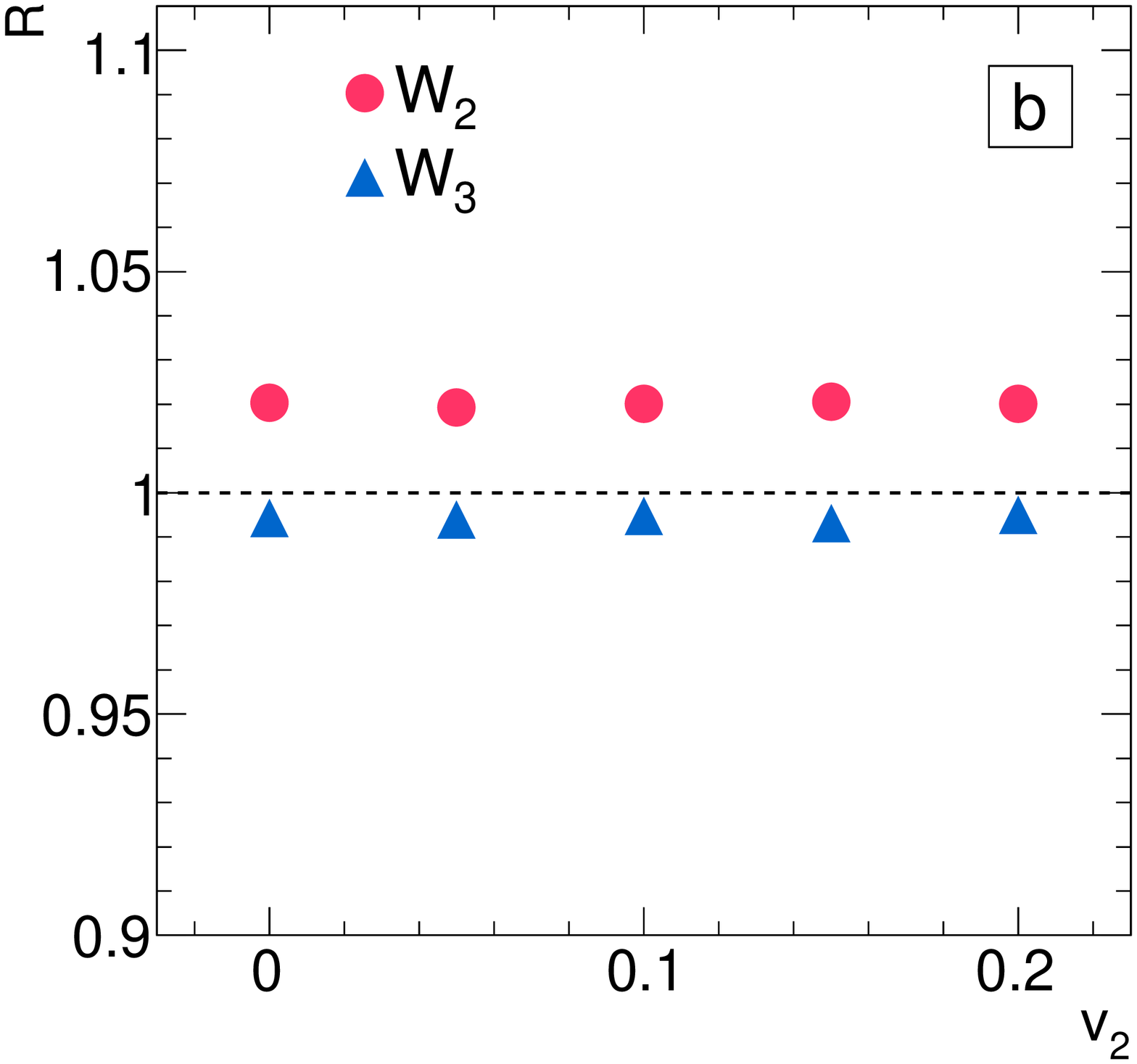} \\
\caption{The toy model simulation on the ratio of root mean square deviation of (a) as a function of $\rho$ resonance $p_{T}$ with fixed value of $v_{\rm 2,\rho}=0.06$ and (b) as a function of $\rho$ resonance flow with fixed value of $p_{\rm T,\rho}=0.6$ GeV/c. Solid line in panel (a) is the expectation value of $W_2$ and $W_3$ with $\delta$ = 0 from analytical calculation, while dash lines represent unity value for eye guidance. All results represent no elliptic flow difference between $\pi^{\pm}$.}
\label{figure 7}
\end{figure}

Figure~\ref{figure 7} shows the ratio of root mean square deviation of numerator and denominator as a function of $\rho$ resonance $p_{\rm T}$ and $v_{2}$, respectively. The elliptic flow difference is set to be zero thus there is background contribution only. In order to investigate the influence of resonance decay angle isolate from other kinematics, we fix $v_{\rm 2,\rho}$ value to be 0.06 in panel (a) and vary the resonance transverse momentum $p_T$ to 0.1, 0.6, 5, 10 and 15 GeV/c, respectively. It can be seen that the ratio is above 1 for $W_3$ (marked concave shape) and below 1 for $W_2$ (represented convex shape) when resonance $p_{\rm T}$ is equal to 0.1 GeV/c. This is because the decay angle between two daughters is close to $\pi$ at low $p_{\rm T}$, and therefore leads to positive (negative) value of covariance to $W_2$ ($W_3$) according to Eq.(\ref{Cov:resonance}) and Eq.(\ref{Cov:resonance(W3)}). When the decay angle is close to $\pi/2$ with increasing $p_{\rm T,\rho}$ to 0.6 GeV/c, the value of $W_2$ is larger than 1 since the covariance becomes negative. On the other hand, the value of $W_3$ is close to 1 because the covariance term of $W_3$ is approximate to zero at the same time. With the increasing of $p_{\rm T,\rho}$ to the extreme case of 15 GeV/c, the open angle between two decay daughters decreases to about zero, leading to both $W_2$ and $W_3$ smaller than 1. Indeed, the expectation value of both $W_2$ and $W_3$ in the case of $\delta = 0$ can be calculated by Eq.(\ref{Ratio}), Eq.(\ref{Cov:resonance}), Eq.(\ref{Var:sigma_W2}), Eq.(\ref{Cov:resonance(W3)}) and Eq.(\ref{Var:sigma_W3}) with a given number of decay particle. It is presented as black solid line in Fig.~\ref{figure 7} and is consistent with the model simulation. Results in panel (b) present $W_2$ and $W_3$ as a function of resonance flow. The $x$ axis is the value of $v_{2,\rho}$. The $v_{\rm 3,\rho}$ is set to be 1/5 of the corresponding $v_{\rm 2,\rho}$. The $p_{\rm T,\rho}$ is 0.6 GeV/c for each point. It is found that both $W_2$ and $W_3$ distributions are flat with the increasing of resonance flow. The trend of $W_2$ and $W_3$ as a function of resonance transverse momentum and flow is consistent with our calculations discussed above.
\begin{figure}
	\captionsetup{justification=raggedright}
	\includegraphics[width=0.8\linewidth]{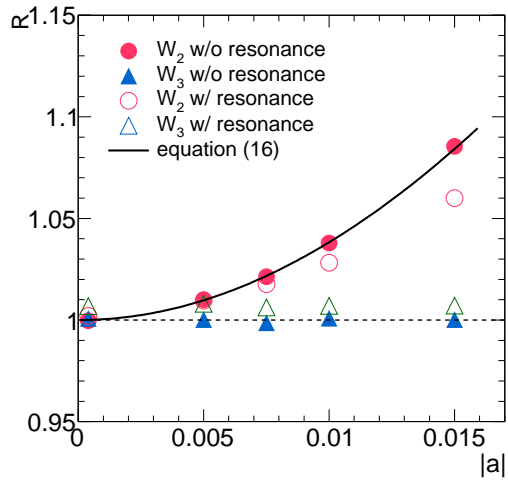}
	\captionof{figure}{The ratio of $\sigma_{\mathrm{num}}/\sigma_{\mathrm{den}}$ as a function of the absolute value of elliptic flow difference. Filled symbols represent the results without resonance decay contribution and open symbols are results with resonance decay taking into consideration in the toy model study. The analytical calculation is illustrated by black solid line.}
	\label{figure 8}
\end{figure}

The response of $W_{\rm n}$ correlator to the CMW effect is presented in Fig. \ref{figure 8}. It can be seen that the filled red circle increases with the increasing of event by event fluctuation elliptic flow difference $a$, and the filled blue triangle remains unity for the studied region. The expected response of $W_2$ to pure signal can be calculated by Eq.(\ref{Var:cmw}) and Eq.(\ref{Var:sigma_W2}) with given particle yield in the event, which is present as solid black curve. On the other hand, the open triangle is slightly above 1 and the open circle is lower than the analytical calculation indicating considerable resonance contribution to our observable. 

\begin{figure*}
\centering
\includegraphics[width=0.3\linewidth]{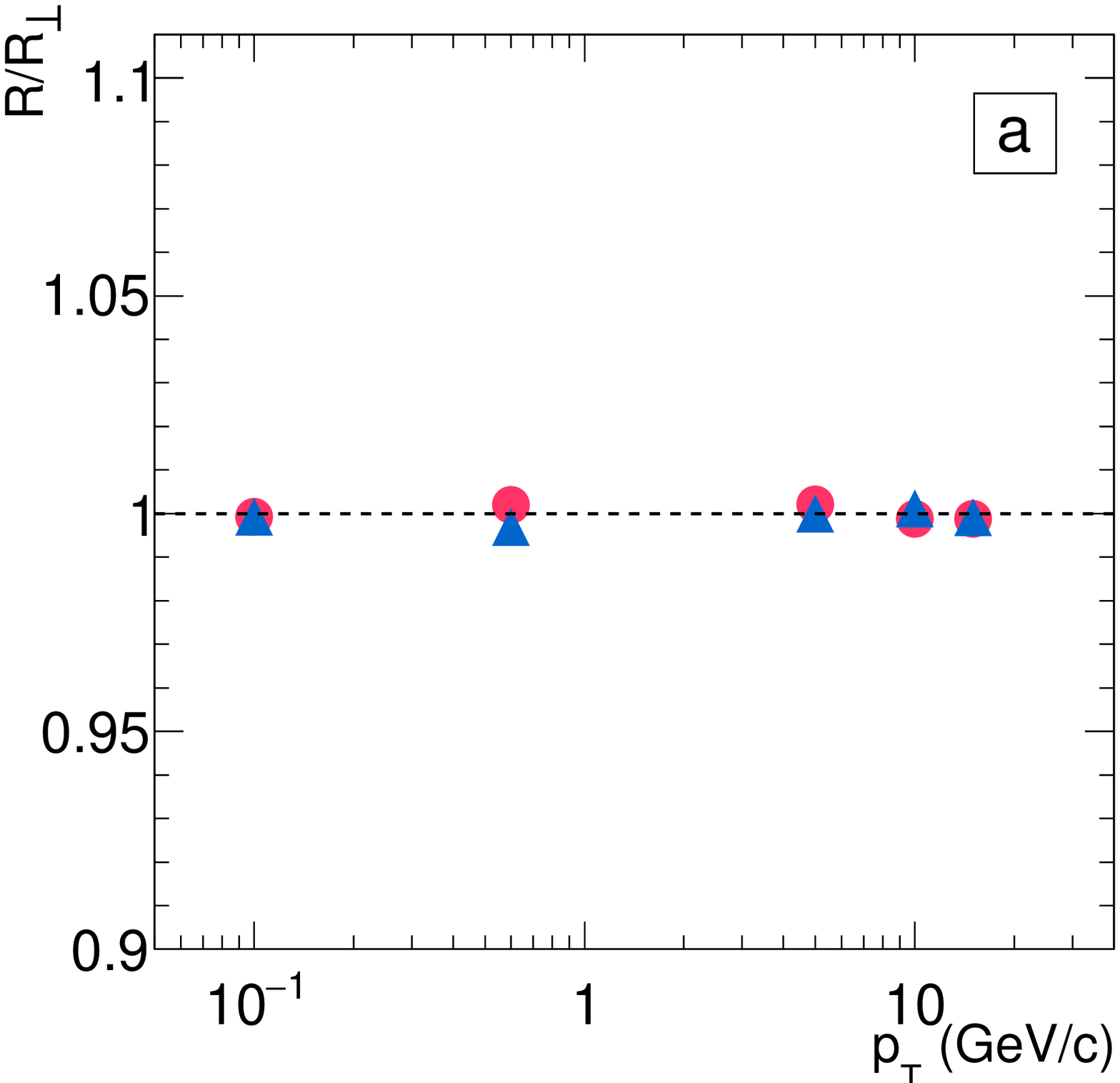}
\includegraphics[width=0.3\linewidth]{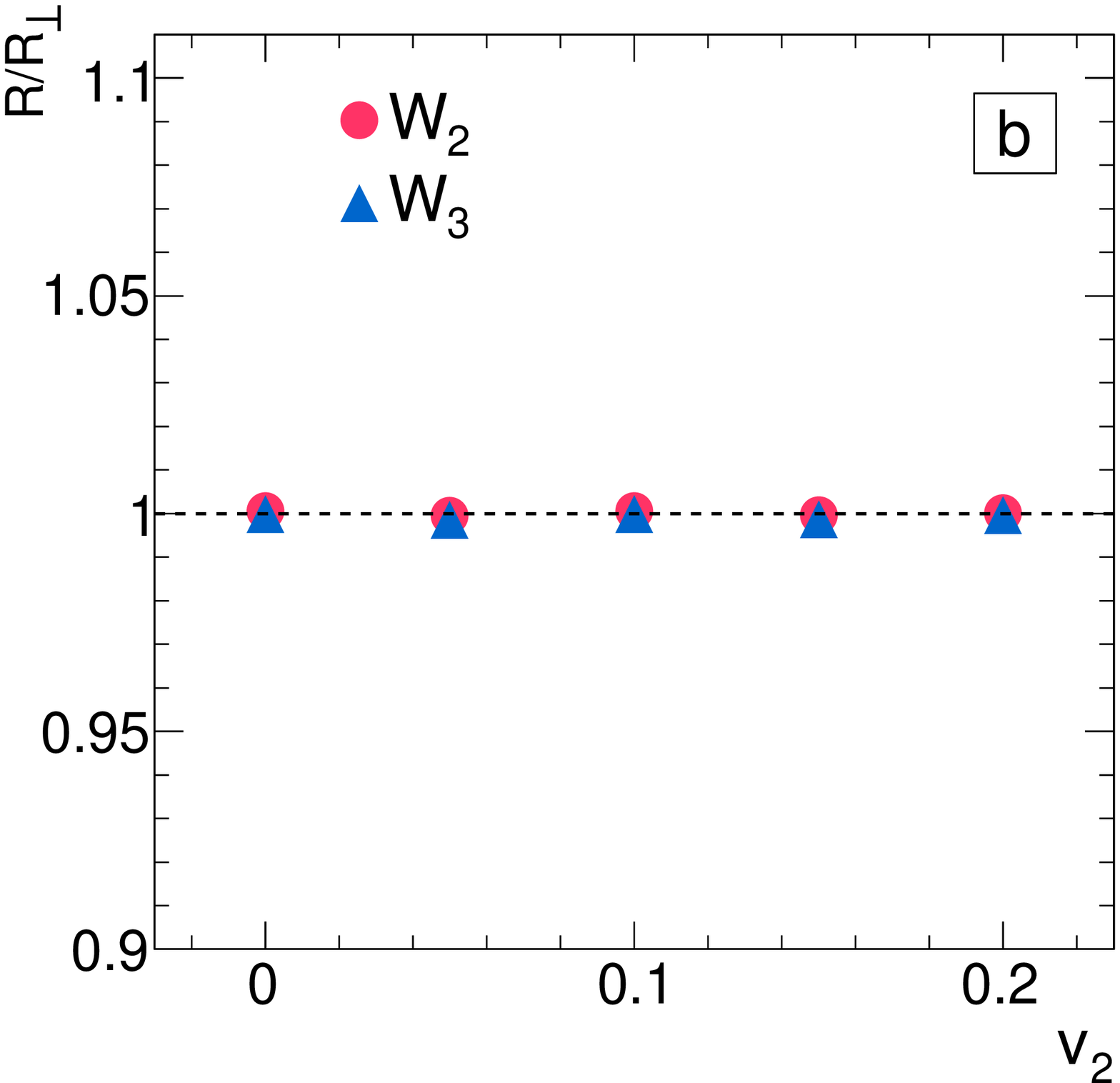}
\includegraphics[width=0.3\linewidth]{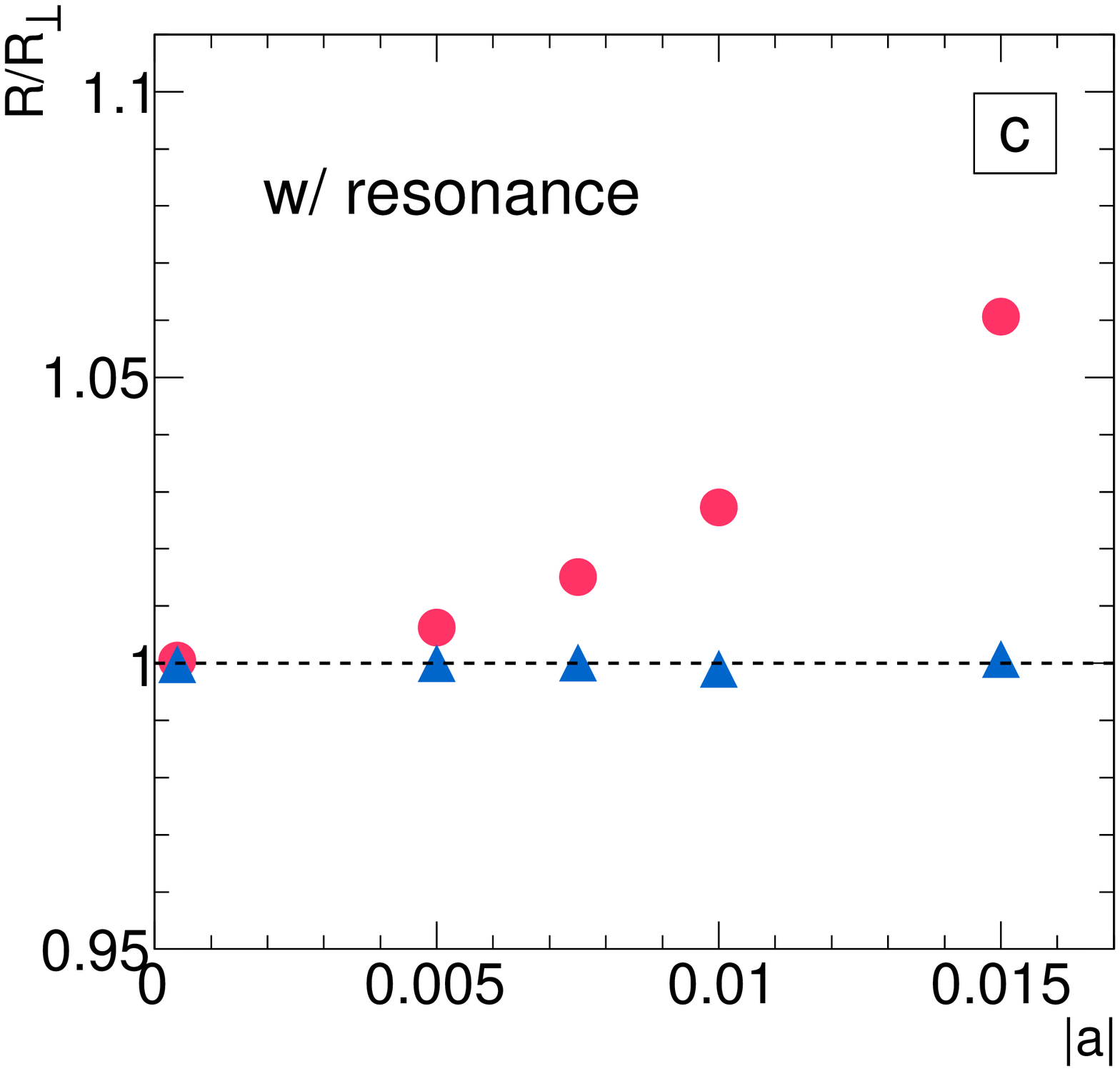}\\
\captionof{figure}{The toy model simulation on the ratio of $R/R_{\rm \perp}$ as a function of (a) resonance $p_{\rm T,\rho}$ with fixed resonance $v_{\rm 2,\rho}=0.06$, (b) resonance flow with fixed $p_{\rm T,\rho}=0.6$ GeV/c, (c) absolute value of elliptic flow difference $\left |a \right |$.}
\label{figure 9}
\end{figure*}

One way to suppress the resonance contribution is to divide the $W_{\rm n}$ correlator by its orthogonal quantities, i.e.
\begin{eqnarray}\label{W:orthogonal}
W_{n\perp} &=& \frac{s_{m}^{+} - s_{m}^{-}}{s_{m}^1 - s_{m}^2} , n=2,3\notag \\
s_m &=& \frac{\sum\limits_{i=0}\limits^{m}sin(n\Delta\phi_{i})}{m}
\end{eqnarray} 
The same analysis procedure has been applied to the $W_{\rm n\perp}$ correlator. It is found that the elliptic flow difference will not contribute to $W_{\rm n\perp}$ but the leading order of resonance contribution remains the same. Similarly, we define $R_{\rm \perp}$ equal to the ratio of root mean square deviation of numerator and denominator of $W_{\rm n\perp}$. Figure~\ref{figure 9} (a) shows the result of the ratio $R/R_{\rm \perp}$ as a function of resonance transverse momentum $p_{\rm T,\rho}$ with fixed $v_{\rm 2,\rho} = 0.06$. The flat distribution has been observed both for $n=2$ and $n=3$. Panel (b) describes the ratio as a function of resonance flow with fixed $p_{\rm T,\rho}$ = 0.6 GeV/c. It is found that the ratio remains unity and insensitive to different flow value. In panel (c), we use the same set-up of resonance spectra and imported flow difference $a$ as described by the open symbols in Figure~\ref{figure 8}. It is seen that the $W_2$ correlator remains sensitive to the event by event fluctuating $v_2$ difference but the resonance contribution has been well suppressed after dividing their corresponding orthogonal quantities.

\section{Summary} \label{sec:sum}
In the framework of the AMPT model with the initial electric quadrupole configuration, we studied the $A_{\rm ch}$ dependence of $v_2$ in 30 - 40\% central Au+Au collisions at $\sqrt{s_{\rm NN}}$ = 200 GeV. With a proper percentage, our model can be used to generate events taking CMW into account and can correctly reproduce the experimental data. It is found that the slope parameters of $\Delta v_2$ on $A_{\rm ch}$ extracted at varied kinematic windows are of the same order of magnitude, and a zero slope is obtained of $\Delta v_3$ on $A_{\rm ch}$. Given that $A_{\rm ch}$ strongly depends on the experimental acceptance and efficiency, we proposed a novel correlator $W$ which specially focuses on the CMW-induced $\Delta v_2$ and is irrelevant to $A_{\rm ch}$. The dependence of the second order correlator $W_2$ on $\Delta v_2$ displays a convex structure in the absence of the quadrupole and a concave shape in the presence of the quadrupole. The curvature of concave structure is related to the strength of the quadrupole and the minimum of $W_2$ is connected to the content and ratio of two quadrupole configurations. The third order correlator $W_3$ serving as a baseline is always consistent with unity. The time evolution of $W$ in AMPT model is also discussed, in particular, the visible effect of final state interactions could weaken the signal. A precise and comprehensive investigation of the resonance background effect on such a correlator is also studied with a toy model. It is found that the resonance contribution can be suppressed by dividing $W_{\rm n}$ by its orthogonal quantities. To sum up, we provides a new method to effectively detect the electric quadrupole moment created by the CMW. Such a method can be performed in the experiment for the further study. 

\section*{Acknowledgements}
We thank Mr Hai Wang for the insight discussion and Dr. Aihong Tang for sharing us with the toy model code for background study. This work was supported in part by the Key Research Program of the Chinese Academy of Science with Grant No. XDPB09, and by the National Natural Science Foundation of China under Contract Nos. 11890714, 11775288, 11605070, 11421505 and 11520101004.

{}

\end{document}